# Aggressive actions and anger detection from multiple modalities using Kinect

AMOL PATWARDHAN, Louisiana State University
GERALD KNAPP, Louisiana State University

Prison facilities, mental correctional institutions, sports bars and places of public protest are prone to sudden violence and conflicts. Surveillance systems play an important role in mitigation of hostile behavior and improvement of security by detecting such provocative and aggressive activities. This research proposed using automatic aggressive behavior and anger detection to improve the effectiveness of the surveillance systems. An emotion and aggression aware component will make the surveillance system highly responsive and capable of alerting the security guards in real time. This research proposed facial expression, head, hand and body movement and speech tracking for detecting anger and aggressive actions. Recognition was achieved using support vector machines and rule based features. The multimodal affect recognition precision rate for anger improved by 15.2% and recall rate improved by 11.7% when behavioral rule based features were used in aggressive action detection.

General Terms: Emotion, Affective computing

Additional Key Words and Phrases: Aggression, multimodal anger recognition, Kinect

## 1. INTRODUCTION

Surveillance systems can highly benefit from automated detection of aggressive activity and anger. An effective and highly responsive security system can be used in several violence and hostility prone places such as prison facilities, mental correctional institutes, sports events and social public gatherings. Early detection of aggressive and provocative actions will help security guards to take measures to prevent the hostile behavior from getting out of control. An affect aware surveillance system can also be useful in detecting aggressive behavior displayed by an agitated customer at a store or hotel counter, identifying angry passengers in a flight or detection of bullying in a school hallway, gymnasium or a sport field. This research used infrared based sensor such as Microsoft Kinect to detect human aggressive activity and anger from multiple modalities such as facial expression, body posture, head movement, hand movement and speech. Research has shown [Castellano et al. 2008] that multimodal affect recognition has better detection accuracy as compared to unimodal affect recognition. According to a study [Gunes and Piccardi 2009] use of temporal features in addition to location based features improved the recognition accuracy. But the researchers identified the lack of emotion recognition systems that have combined data from head, face, hand, body and speech, all at once. Hence the first objective of this work was to study the accuracy, precision and recall rates of aggressive activity and anger detection from head, face, body, hand and speech data channels, unified in the same emotion recognition system. The feature vector used for training the classifiers was obtained from position and temporal data across multiple frames. The second objective of the research was studying the use of rule based features to boost the recognition rates of the automatic aggression recognition process. The feature vector for the rule based features contained motion data from face, limbs and body across multiple frames.

## 2. LITERATURE REVIEW

Researchers used dynamic Bayesian networks [Zajdel et al. 2007] for multimodal fusion and studied the use of low level descriptors of environmental noise and scene such as screams, passing of train and its use in aggressive human behavior. In a research work [Lefter et al. 2012] information about feature correlation obtained during fusion processes was used to improve aggression detection prediction. Violent scenes from videos were detected using multi-step process in a study [Giannakopoulos et al. 2010]. In this research the initial step estimated probabilistic measures of classes and the second step used ensemble classification for fusion of audio visual data. Researchers [Datta et al. 2002] used motion trajectory and orientation of limbs for detecting violent human activity based on acceleration measure vector.

A study done on urban environments [Andersson et al. 2010] used two steps process for aggressive behavior recognition. The first step used low level event detection from audio and video channels and the second step combined the data from individual modalities to perform disambiguation. Results indicated degradation of performance when audio channel was excluded or the number of video cameras was reduced. In a study [Mecocci and Micheli 2007] low level feature such as spatial-temporal behavior of color stains, motion parameters and warping was analyzed to predict aggressive human activity based on maximum warping energy index. A research work [Metallinou et al. 2008] showed that the classifier combinations in multimodal emotion recognition resulted in higher performance. The study used Bayesian classifier weighting scheme and support vector machines that used post classification accuracies as the features. Gaussian mixture models were used for individual modalities. A study [Schuller et al. 2002] analyzed multiple modalities such as speech and haptical interation to detect user emotion. The features used for speech modality were pitch, prosody, word rate and degree of verbosity.

Research work [De Silva 2004] showed that the combination of audio and visual features using a rule based technique improved the emotion recognition accuracy for the six universal emotions. Researchers [Zeng et al. 2008] used multistream fused hidden markov model to develop an algorithm for detecting four different affective states from audio and visual modalities. A study [Sung et al. 2011] performed on human activity recognition used hierarchical maximum entropy model. In this study smaller subsets of human activity were used to construct a two-layer graph using a dynamic programming approach. A survey [Aggarwal and Ryoo 2011] provided human activity recognition methodologies and public datasets. The survey discussed performance comparison, statistical approach, syntactical and descriptive approach for a single person activity analysis.

Researchers [Cheng et al. 2011] focused on small human groups to study the human activity based on motion and appearance based spatiotemporal features. The study used multiple kernel learning methods to recognize group activities. A research [Zhang and Sawchuk 2012] provided a statistical framework to model human activity based on primitive motion based features. The study described a bag of features extracted from a stream of continuous sensor data by dividing into smaller relevant windows. A study [Ghasem-Aghaee et al. 2009] used fuzzy expert system to simulate anger and aggression in an intelligent agent. In a study mid-level predictions were fused using low level feature data, extracted from fight sequences in Hollywood movies to detect violence. A study [Zhang et al. 2013] provided answers to challenges in communication and interaction between humans and computers that involved multiple modalities.

## 3. METHOD OVERVIEW

In this research we proposed detection of aggressive actions and anger using multiple modalities. The detection was achieved using two stage process involving anger and aggression recognition classification with first stage using support vector machine and the second stage using rule based features. The feature vector definition for both stages is described in the following sections. The experimental setup included a Kinect sensor which provided multimodal input data to the computer system running the affect recognition program. Joint location and movement, facial expressions and words were tracked from multiple modalities such as hand movement, body posture, head gesture, face and speech.

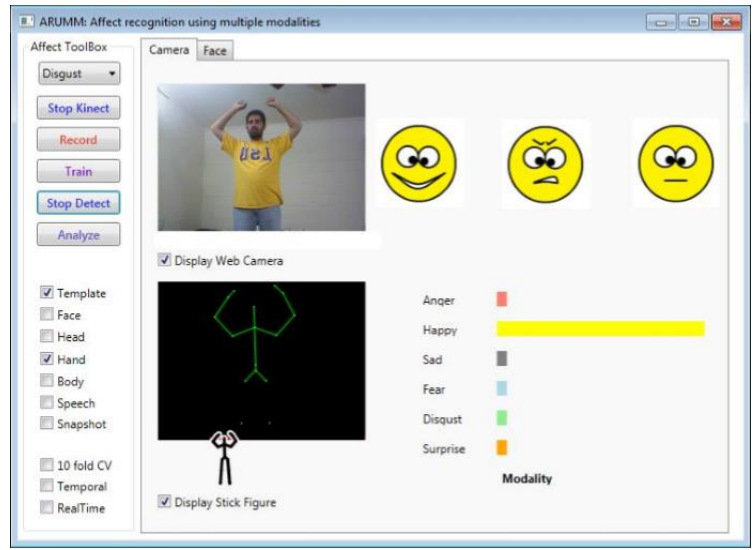

Fig 1. Affect recognition using multiple modalities.

6 basic emotions [Ekman, 1999] and 1 neutral state were used to define class labels for a total of 7 classes. Five individuals enacted aggressive actions, angry facial expressions and other non-aggressive emotional gestures for 21 sessions (3 acts for each of the 7 emotional state). Duration of each act varied between 15 seconds to 1 minute. This resulted in a total of 105 (5 x 7 x 3) sessions. The individuals also used speech, raised voice and tone to express anger and aggression. The data was captured in controlled lighting with full frontal orientation with a distance of 1.5 to 4 meters from the Kinect infrared sensor camera. The dataset was split into 80% training data and 20% test data. The classifier was trained using 10- fold cross validation. Radial basis function was used as the non-linear kernel function. The slack variable C was set to 1 and gamma value for head, face, body and hand modality were 1/12, 1/60, 1/8 and 1/16 respectively. Finally, the detection results were evaluated on the 20% test data which was not used in the initial training of the classifiers. We implemented the multimodal emotion recognition system using C#.NET and Kinect API and integrated it with speech based emotion detection library called openEar [Eyben et al. 2009] and called it ARUMM (Affect recognition using multiple modalities).

The list of actions was developed based on ideas from existing research [Gunes and Piccardi, 2009], [Castellanos, et al. 2008]. The authors explained the situation to the actors and the actors reacted with one of the actions and dialog from the available list. The actors were given freedom to spontaneously enact actions and dialogs from the list or even improvise.

Table I. Description of gestures for dataset generation

| Modality | Gesture |
|---|---|
| Anger | Stand with arms on hips. |

|  | Stand with arms folded. |
|---|---|
|  | Stand using a stance for fist fight. |
|  | Move forward with arms raised in anger. |
|  | Action of throwing an object in anger. |
|  | Punching. |
|  | Holding head in frustration. |
|  | Throwing a fit. |
|  | Raising arms in rage. |
|  | Moving around aggressively. |
|  | Pointing a finger at someone. |
|  | Threaten someone. |
|  | Scowl. |
| Happy | Jump with joy. |
|  | Express joy with arms close to chest. |
|  | Raise arms in joy. |
|  | Laughing. |
|  | Smiling. |
|  | Fist pumping in joy. |
| Sad | Hold head with one hand in despair. |
|  | Move around with slouched back. |
|  | Hold hands on the face. |
|  | Hold head with hands. |
|  | Looking down leaning against wall. |
|  | Looking down with hands on waist. |
|  | Looking down with hands folded. |
|  | Crying. |
|  | Tooth ache. |
|  | Head hurt. |
| Surprise | Express surprise with arms close to chest. |
|  | Express surprise with one arm on chest. |
|  | Express surprise with arms on the side slightly away from hips. |
|  | Raising arms in surprise. |
|  | Moving back in surprise. |
|  | Walking forward and getting startled. |
|  | Holding arms near chest in surprise. |
|  | Covering mouth with hands. |
| Fear | Action of moving away in fear. |
|  | Hold arms near chest and express fear. |
|  | Duck. |
|  | Action of moving to the side. |
|  | Moving backwards trying to evade. |
|  | Moving sideways. |
|  | Looking up and run away. |
|  | Getting rid of an insect on shirt. |
| Disgust | Fold arms and shrug with shoulder. |
|  | Move away in one direction. |
|  | Hold nose with one arm and evade with other. |
|  | Moving side-ways with arms evading. |
|  | Looking down expressing disgust. |
|  | Moving back in disgust. |
| Neutral | Stand in neutral position in front of Kinect. |

### 3.1 Feature Definition

The location based feature vector used for hand movement was defined by co-ordinates of the shoulder, elbow and wrist joints of both arms. The angle made by each pair of joints with the horizontal axis was also used as a feature. In addition to the location based feature a temporal feature vector was also used to train the classifier. The temporal feature vector was defined as the velocity of the movement of the joints across two consecutive frames. The x and y component of the velocity vector was calculated using the displacement in the x and y co-ordinates of the joints divided by the time difference between the consecutive frames. The location based and temporal feature vector are shown in Equation (1) and Equation (2) respectively.

$$FVL_{hand} = \{Pt_1(x), Pt_1(y), \ldots, Pt_n(y), \theta_1(x,y), \ldots, \theta_m(x,y)\} \quad (1)$$

where n is the total number of joints tracked and m is the total number of pair of joints. $Pt_i(x)$ is the x-coordinate of the $i^{th}$ tracked point. The suffix L is used to indicate feature vector is based on location of joints.

$$FVT_{hand} = \{\Delta Vt_1(s), \Delta Vt_2(s) \ldots, \Delta Vt_n(s), \theta_1(x,y), \ldots, \theta_n(x,y)\} \quad (2)$$

where n is the total number of joints tracked and $\Delta Vt_i(s)$ is the magnitude component of the velocity vector of the $i^{th}$ displacement. $\theta_i$ is the orientation of the velocity vector. The suffix T is used to indicate that the feature vector is based on movement of joints. The feature vector for the head modality was defined using 12 tracked points on the head as shown in Equation (3) and (4).

$$FVL_{head} = \{Pt_1(x), Pt_1(y), \ldots, Pt_n(y), \theta_1(x,y), \ldots, \theta_m(x,y)\} \quad (3)$$

where n = 12 for location and temporal feature vector for head modality.

$$FVT_{head} = \{\Delta Vt_1(s), \Delta Vt_2(s) \ldots, \Delta Vt_n(s), \theta_1(x,y), \ldots, \theta_n(x,y)\} \quad (4)$$

The location and temporal feature vector for the facial expression were defined using 60 tracked points from the facial region as shown in Equation (5) and Equation (6).

$$FVL_{face} = \{Pt_1(x), Pt_1(y), \ldots, Pt_n(y), \theta_1(x,y), \ldots, \theta_m(x,y)\} \quad (5)$$

where n = 60 for location and temporal feature vector from face modality.

$$FVT_{face} = \{\Delta Vt_1(s), \Delta Vt_2(s), \ldots, \Delta Vt_n(s), \theta_1(x,y), \ldots, \theta_n(x,y)\} \quad (6)$$

The 60 tracked points on the facial region included points such as lower and upper lips, lower and upper eye lid, left and right cheek, chin, tip of nose etc. The location and temporal feature vector definition for body posture modality is provided in Equation (7) and Equation (8).

$$FVL_{body} = \{Pt_1(x), Pt_1(y), \ldots, Pt_n(y), \theta_1(x,y), \ldots, \theta_m(x,y)\} \quad (7)$$

where n = 12 for location and temporal feature vector from body posture modality.

$$FVT_{body} = \{\Delta Vt_1(s), \Delta Vt_2(s), \ldots, \Delta Vt_n(s), \theta_1(x,y), \ldots, \theta_n(x,y)\} \quad (8)$$

The speech modality was based on speech recognition application programming interface (API) from Microsoft. The affect recognition implementation in the research used a dictionary of words associated with anger to perform a lookup whenever the user uttered a word. If there was a match from the set of words defining the anger emotion class with a confidence level of more than 0.3 then the detected emotion was assigned the anger class. The dictionary was stored in an xml file. The dictionary used for lookup is described as follows.

$Dictionary_{anger}$ = {Beat, Hate, Kill, Punch, What, Angry, Annoy, Vex, Trash, Smack, Whack, Sock, Fist, Kick, Shove, Push, Hit, Ram, Wrestle, Nerd, Loser, Affronted,

*Belligerent, Bitter, Burned, Enraged, Fuming, Furious, Heated, Incensed, Infuriated, Intense, Outraged, Provoked, Seething, Storming, Truculent, Vengeful, Vindictive, Wild, Aggravated, Annoyed, Antagonistic, Crabby, Cranky, Exasperated, Fuming, Grouchy, Hostile, Ill-tempered, Indignant, Irate, Irritated, Offended, Ratty, Resentful, Sore, Spiteful, Testy, Ticked off, Bugged, Chagrined, Dismayed, Galled, Grim, Impatient, Irked, Petulant, Resentful, Sullen, Uptight, Freak, Emo, Whale, Pig, Fat, Wannabe, Poser, Mad, Crazy, Dead, Damn,…..}*

The dictionary provided above is not the complete list and there are more words in the set that have been excluded due to the expletive nature.

**3.2 Rule based features**

In addition to the existing approaches to feature definition we proposed development of novel features from human behavior and body language. This research proposed a concept of "behavioral rule based features" to estimate an emotion and evaluated their effectiveness in emotion prediction when used along with geometric features for training a classifier. This research proposed a set of rules and temporal features across multiple frames to track movement of each feature point. Each rule was derived from emotional human behavior. For instance, rules were developed to track nodding the head, shrugging the shoulders and raising eyebrows.

The motivation to use rules for emotion estimation was drawn from research done in emotional gesture recognition [Zhang and Yap, 2012] and adaptive rule based facial expression recognitions [Ioannou, et al. 2004]. The studies have demonstrated successful affect recognition by using limited set of gesture based rules and rules extracted from various facial expression profiles. (Coulson, 2010) used 176 computer generated mannequins from shoulder, hand, head descriptors and showed that each posture and movement can be attributed to one of the six basic emotions.

The rule based features for anger and aggressive activity depend on evaluation of a series of rules that define anger and aggression. The number of evaluations that satisfied the rule definition determines the confidence level of the aggression recognition. For instance, if 5 rules denoted by R out of 15 evaluations denoted by N satisfied the criteria that the person has an angry face then the confidence level was calculated as $R/N = 5/15 = 0.33$. The average of confidence level was calculated over a series of 10 continuous frames of tracked data. An illustrative list of the rules defining anger or aggressive activity used by the various modalities is as follows:

(1) If the y-coordinates of the left and right wrist are below respective y-coordinates of elbow and the x-coordinate of the left wrist is less than the x-coordinate of the elbow and x-coordinate of the right wrist is greater than x-coordinate of the right elbow then the rule evaluated to true.
(2) If the y-coordinates of the left and right wrist are above the respective y-coordinate elbow and the y-coordinate of either left or right wrist is above the other then the rule evaluated to true.
(3) If the x-coordinate of the left and right wrist moved to left or right by a distance of 15px then the rule evaluates to true.
(4) If the y-coordinate of the left and right wrist decreased by a distance of 15px then the rule evaluated to true.
(5) If the center of the head moved left and right by a distance of 15px for 3 or more times, then the rule evaluated to true.

(6) If the distance between the x-coordinates of top of right eyebrow and the top of left eyebrow is less than the distance between the x-coordinates of top of left and right upper lip then the rule evaluated to true.
(7) If the difference between the distance between the x-coordinates of the left and right feature points on the upper lip and the distance between x-coordinate of the left and right feature point on the lower lip, was within 5px then the rule evaluated to true.
(8) If the distance between top of left and right eyebrow and the top of left and right eyelid was greater than the distance between top of left and right eyelid and the top of left and right corner of the eye then the rule evaluated to true.
(9) If the y-coordinate of the bottom of the chin moved up and down 3 times by a distance of 5px or more, then the rule evaluated to true.
(10) If the y-coordinate of the top center of the head moved up by 15px or more and the x-coordinate of the top center of the head moved left and right by 3px or more for 3 or more times, then the rule evaluated to true.

The Rule (1) indicates that the user is standing with hands on the waist. The Rule (2) indicates that the person has hands in a position to fist fight. Similarly, the rest of the rules define the actions such as an angry facial expression, raising arms to get ready to fight, frowning, sneering or shaking the head in disapproval, in terms of movement of the tracked feature points in x, y direction and the number of times the feature point moves and whether the movement is above or below a threshold distance and count.

### 3.3 Fusion Strategy

The research used decision level fusion based on majority voting. Based on the classifier outcome and the rule based feature classifier outcome each modality detected the current user as angry and aggressive or not angry. The fusion process was event driven. Every time a modality completed anger recognition an emotion detected event was raised which was handled by the fusion processing function. In this process the outcome of anger and aggression recognition from various modalities was used to determine whether anger class had received the highest vote from all the available outcomes at a given instance of time.

Only the outcome above confidence level threshold T was considered. Based on experimental results the threshold of T >= 0.218 provided the best precision and recall rate results.

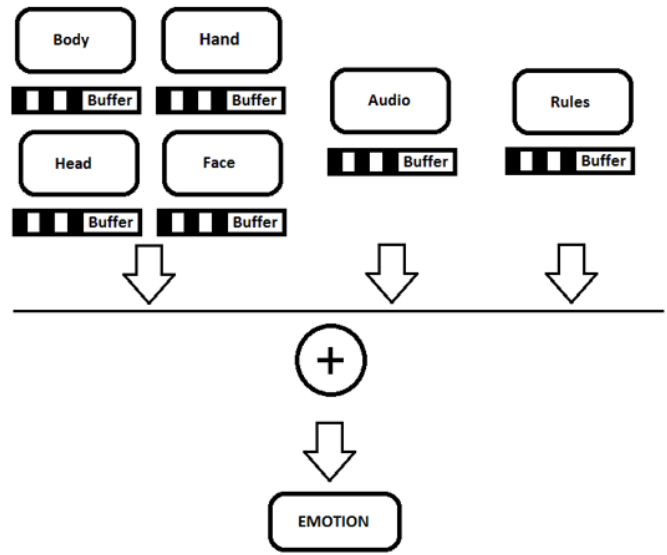

Fig 2. Decision level fusion for multimodal emotion recognition.

Once the frame level outcome was determined from various modalities, it was then used to determine the recognition outcome of the session. For instance, the affect recognition outcome for a session of 1 minute, was determined based on the number of time anger was detected per total number of fusion execution runs. For example, if anger was detected 37 times out of 50 multimodal fusions in the span of one minute then the session confidence level for anger and aggressive behavior was 0.74.

## 4. RESULTS

Each individual participating in the study was asked to enact anger or aggression in front of the Kinect sensor. The participants also used aggressive language to express their anger. The recognition of aggressive behavior was done using only position and temporal feature based classifiers for the first set of results as discussed in the following section.

### 4.1 SVM Based Results

The overall performance of classifier for multiple modality using position and temporal features was 78.8%. The emotions with high precision were fear (83.3%) and surprise (81.7%). The precision for anger (74.4%), disgust (73.2%) and happy (76.8%) was low compared to fear and surprise.

Table II. Multiple Modality

| Emotion | Precision | Recall | F-Measure | Class |
|---|---|---|---|---|
| Anger | 0.744 | 0.8 | 0.77 | 0 |
| Happy | 0.768 | 0.736 | 0.751 | 1 |
| Surprise | 0.817 | 0.849 | 0.832 | 2 |
| Disgust | 0.732 | 0.74 | 0.735 | 3 |
| Fear | 0.833 | 0.759 | 0.794 | 4 |
| Sad | 0.784 | 0.795 | 0.789 | 5 |
| Neutral | 0.829 | 0.838 | 0.833 | 6 |

The following results show the confusion matrix of classifier for multiple modality using position and temporal geometric features.

Table III. Multiple Modality Confusion Matrix

| Emotion | Anger | Happy | Surprise | Disgust | Fear | Sad | Neutral |
|---|---|---|---|---|---|---|---|
| Anger | 0.8 | 0.051 | 0.04 | 0.062 | 0 | 0.047 | 0 |
| Happy | 0.081 | 0.736 | 0.055 | 0.033 | 0.021 | 0.021 | 0.054 |
| Surprise | 0.028 | 0.048 | 0.849 | 0 | 0.055 | 0.018 | 0.002 |
| Disgust | 0.104 | 0.053 | 0.002 | 0.74 | 0.036 | 0.06 | 0.004 |
| Fear | 0.011 | 0.04 | 0.024 | 0.037 | 0.759 | 0.059 | 0.069 |
| Sad | 0.05 | 0.023 | 0 | 0.067 | 0.029 | 0.795 | 0.036 |
| Neutral | 0.004 | 0.025 | 0.061 | 0.031 | 0.038 | 0.004 | 0.838 |

The most recognized emotions were surprise (83.8%) and anger (80%). Happy (73.6%), disgust (74%) and fear (75.9%) were recognized at a very similar level of 74%. Surprise was misclassified with fear (5.5%) and happy (4.8%). Anger was misclassified with happy (5.1%) and disgust (6.2%).

### 4.2 SVM + Rule Based Results

The study proposed use of rule based anger prediction to augment the results from the classifier based recognition. The results of the experiment are shown in the following section. The overall performance of classifier for multiple modality using rule based features was 93.51%. The emotions with high precision were anger (99.6%), sad (99.4%), disgust (98.3%) and happy (91.1%). The precision for surprise (86.1%) and fear (85.5%) was low compared to the other 4 emotions.

Table IV. Multimodal Emotion Recognition (Rule Based Features)

| Emotion | Precision | Recall | F-Measure | Class |
|---|---|---|---|---|
| Anger | 0.996 | 0.917 | 0.955 | 0 |
| Happy | 0.911 | 0.858 | 0.883 | 1 |
| Surprise | 0.861 | 0.992 | 0.922 | 2 |
| Disgust | 0.983 | 0.996 | 0.990 | 3 |
| Fear | 0.855 | 0.988 | 0.916 | 4 |
| Sad | 0.994 | 0.838 | 0.909 | 5 |
| Neutral | 1.000 | 0.973 | 0.986 | 6 |

The following results show the confusion matrix of classifier for multiple modality using rule based features. The most recognized emotions were disgust (99.6%), surprise (99.2%), fear (98.8%) and anger (91.7%). Sad was misclassified with fear (16.2%). Happy was misclassified with surprise (13.9%). The misclassifications occurred for emotions within the same emotion quadrant on valence-arousal axis.

Table V. Multimodal (Rule Based) Confusion Matrix

| Emotion | Anger | Happy | Surprise | Disgust | Fear | Sad | Neutral |
|---|---|---|---|---|---|---|---|
| Anger | 0.917 | 0.073 | 0.006 | 0.003 | 0.001 | 0 | 0 |
| Happy | 0.001 | 0.858 | 0.139 | 0.001 | 0 | 0 | 0 |
| Surprise | 0 | 0.008 | 0.992 | 0 | 0 | 0 | 0 |
| Disgust | 0 | 0 | 0.003 | 0.996 | 0 | 0 | 0 |
| Fear | 0 | 0 | 0 | 0.009 | 0.988 | 0.003 | 0 |
| Sad | 0 | 0 | 0 | 0 | 0.162 | 0.838 | 0 |
| Neutral | 0.003 | 0.006 | 0.005 | 0.005 | 0.004 | 0.003 | 0.973 |

The classification, precision and recall rates showed improvement by combining the classifier and rule based recognition. The rule based features were not applicable to the speech modality because recognition was done using dictionary lookup and did not include features that could be used in rule based evaluation of emotion.

## 5. CONCLUSIONS

In this research, we implemented position and temporal feature based aggression and anger detection using data obtained from infrared sensor such as Microsoft Kinect. Multiple modalities were used for tracking various facial points, body and limb joints and speech. The combination of supervised learning classification and rule based features for determination of anger and aggressive activity showed improvement in results. The fusion strategy used during multimodal affect recognition proved efficient because of the ability to accommodate absence of outcomes from individual modalities at a given instance of time. The precision and recall for anger recognition were in the range of 74.4% and 80% respectively. The precision improved to 99.6% and recall improved to 91.7%. Thus the behavioral rule based features improved the anger recognition performance. The results indicated that an infrared equipped surveillance system is indeed feasible and can be used in early detection of aggressive behavior. As a future scope the study would consider spontaneous aggressive behavior in a real environment instead of enacted anger in a controlled setting of the experimental setup.

**REFERENCES**

Jake . K. Aggarwal and Michael. S. Ryoo. 2011. Human activity analysis: A review. ACM Comput. Surv. 43, 3, Article 16 (April 2011), 43 pages. DOI=10.1145/1922649.1922653 http://doi.acm.org/10.1145/1922649.1922653.

Maria Andersson, Stavros Ntalampiras, Todor Ganchev, Joakim Rydell, Jörgen Ahlberg, Nikos Fakotakis. 2010. Fusion of acoustic and optical sensor data for automatic fight detection in urban environments, In *Proceedings of the 13th Conference on Information Fusion (FUSION'10)*, 1-8.

Ginevra Castellano, Loic Kessous, and George Caridakis. 2008. Emotion Recognition through Multiple Modalities : Face , Body Gesture , Speech. *Emotion,* Vol. 4868, 92-103.

Zhongwei Cheng, Lei Qin, Qingming Huang, Shuqiang Jiang, Shuicheng Yan, and Qi Tian. 2011. Human group activity analysis with fusion of motion and appearance information. In *Proceedings of the 19th ACM international conference on Multimedia (MM '11)*. ACM, New York, NY, USA, 1401-1404. DOI=10.1145/2072298.2072025 http://doi.acm.org/10.1145/2072298.2072025.

Liyanage C. De Silva. 2004. Audiovisual emotion recognition, In *Proceedings of the IEEE International Conference on Systems, Man and Cybernetics*, Vol. 1, 649-654.

Ankur Datta , Mubarak Shah , Niels Da , Niels Da Vitoria Lobo. 2002. Person-on-person violence detection in video data. In *Proceedings of the International Conference on Pattern Recognition*, vol. 1, 433–438.

Nasser Ghasem-Aghaee, Bardia Khalesi, Mohammad Kazemifard, and Tuncer I. Ören. 2009. Anger and aggressive behavior in agent simulation. In *Proceedings of the 2009 Summer Computer Simulation Conference (SCSC '09). Society for Modeling & Simulation International*, Vista, CA, 267-274.

A. S. Patwardhan, 2016. "Structured Unit Testable Templated Code for Efficient Code Review Process", *PeerJ Computer Science*.

A. S. Patwardhan, and R. S. Patwardhan, "XML Entity Architecture for Efficient Software Integration", International Journal for Research in Applied Science and Engineering Technology (IJRASET), vol. 4, no. 6, June 2016.

A. S. Patwardhan and G. M. Knapp, "Affect Intensity Estimation Using Multiple Modalities," Florida Artificial Intelligence Research Society Conference, May. 2014.

A. S. Patwardhan, R. S. Patwardhan, and S. S. Vartak, "Self-Contained Cross-Cutting Pipeline Software Architecture," International Research Journal of Engineering and Technology (IRJET), vol. 3, no. 5, May. 2016.

Theodoros Giannakopoulos, Alexandros Makris, Dimitrios I. Kosmopoulos, Stavros J. Perantonis, Sergios Theodoridis. 2010. Audio-Visual Fusion for Detecting Violent Scenes in Videos. In *Proceedings of the Hellenic Conference on Artificial Intelligence*, 91–100.

Hatice Gunes, and Massimo Piccardi. 2009. Automatic Temporal Segment Detection and Affect Recognition from Face and Body Display. *IEEE Transactions on Systems, Man, and Cybernetics – Part B*, 39, 64-84.

Bogdan Ionescu, Jan Schlüter, Ionut Mironica, and Markus Schedl. 2013. A naive mid-level concept-based fusion approach to violence detection in Hollywood movies. In *Proceedings of the 3rd ACM conference on International conference on multimedia retrieval (ICMR '13)*. ACM, New York, NY, USA, 215-222. DOI=10.1145/2461466.2461502 http://doi.acm.org/10.1145/2461466.2461502.

Iulia Lefter, Gertjan J. Burghouts, Léon J. M. Rothkrantz. 2012. Automatic Audio-Visual Fusion for

Aggressive actions and anger detection from multiple modalities using Kinect                                    39:11Aggression Detection Using Meta-information. In *Proceedings of the Advanced Video and Signal Based Surveillance (AVSS '12)*, 19–24.

Alessandro Mecocci, Francesco Micheli. 2007. Real-Time Automatic Detection of Violent-Acts by Low-Level Colour Visual Cues, In *Proceedings of IEEE International Conference on Image Processing*, Vol. 1, 345-348.

Angeliki Metallinou, Sungbok Lee and Shrikanth S. Narayanan. 2008. Audio-visual emotion recognition using Gaussian mixture models for face and voice, In *Proceedings of the IEEE International Symposium on Multimedia (ISM)*, Berkeley, USA.

Björn Schuller, Manfred Lang, Gerhard Rigoll. 2002. Multimodal emotion recognition in audiovisual communication, In *Proceedings of the IEEE International Conference on Multimedia and Expo (ICME'02)*, Vol. 1, 745–748.

Eyben, F., Wollmer, M., and Schuller, B. 2009. openEAR - Introducing the Munich Open-Source Emotion and Affect Recognition Toolkit. *Proceedings of the 4th International HUMAINE Association Conference on Affective Computing and Intelligent Interaction 2009 (ACII 2009)*, 1-6. doi:10.1109/ACII.2009.5349350.

Jaeyong Sung, Colin Ponce, Bart Selman, Ashutosh Saxena. 2011. Human Activity Detection from RGBD Images, In *Proceedings of the AAAI workshop on Pattern, Activity and Intent Recognition (PAIR)*.

Wojtek Zajdel, Johannes D. Krijnders, Tjeerd C. Andringa, Dariu M. Gavrila. 2007. CASSANDRA: audio-video sensor fusion for aggression detection. In *Proceedings of the Advanced Video and Signal Based Surveillance (AVSS '07)*.

Zhihong Zeng, Jilin Tu, Brian Pianfetti, Thomas S. Huang. 2008. Audio–Visual Affective Expression Recognition Through Multistream Fused HMM, *IEEE Transactions on Multimedia*, Vol. 10, 570-577.

Mi Zhang and Alexander A. Sawchuk. 2012. Motion primitive-based human activity recognition using a bag-of-features approach. In *Proceedings of the 2nd ACM SIGHIT International Health Informatics Symposium (IHI '12)*. ACM, New York, NY, USA, 631-640. DOI=10.1145/2110363.2110433 http://doi.acm.org/10.1145/2110363.2110433.

Yang Zhang, Li Zhang, and Alamgir Hossain. 2013. Multimodal intelligent affect detection with Kinect: extended abstract. In *Proceedings of the 2013 international conference on Autonomous agents and multi-agent systems (AAMAS '13)*. International Foundation for Autonomous Agents and Multiagent Systems, Richland, SC, 1461-1462.

Ekman, P. (1999). Basic emotions. *Cognition, 98*(1992), 45-60.

Coulson, M. 2010. Attributing emotion to static body postures: recognition accuracy, confusions, and viewpoint dependence. *Journal of Nonverbal Behavior*, 28(2), 117-139.

A. S. Patwardhan, "An Architecture for Adaptive Real Time Communication with Embedded Devices," LSU, 2006.

A. S. Patwardhan and G. M. Knapp, "Multimodal Affect Analysis for Product Feedback Assessment," IIE Annual Conference. Proceedings. Institute of Industrial Engineers-Publisher, 2013.

Zhang, L., & Yap, B. 2012. Affect Detection from Text-Based Virtual Improvisation and Emotional Gesture Recognition. Advances in Human-Computer Interaction, 2012(461247). doi:10.1155/2012/461247.

Ioannou, S., Raouzaiou, A., Karpouzis, K., Pertselakis, M., Tsapatsoulis, N., & Kollias, S. (2004). Adaptive rule-based facial expression recognition. Lecture Notes in Artificial Intelligence, 3025, 466-475.

A. S. Patwardhan and G. M. Knapp, "EmoFit: Affect Monitoring System for Sedentary Jobs," preprint, arXiv.org, 2016.